\documentclass {article}
\usepackage{hyperref} \usepackage{url}
\def\nwph {Newtonian physics}
\def \tf  {time-function}
\def \tfunction {time-function}
\def\mlr {Marc Lachi\`eze-Rey}
\def\lrm {Lachi\`eze-Rey, M.}
\def\spt {space-time}

\def \Eequs  {Einstein equations}
\def \wline{ world line}
\def\cmb {Cosmic Microwave Background}
\def\coord {coordinate}

\def\calA {{\cal A}}
 
\def\calB {{\cal B}}

\def\calO {{\cal O}}

\def\guill {\textquotedblleft ~}
\def\wrt  {w.r.t.~}\def\qgr  {quantum gravity}

\newcommand{\arxiv} [1] { \url{http://arxiv.org/abs/#1} }  
\newcommand{\arXiv} [1] { \url{http://arxiv.org/abs/#1} }

\def\Mink {Minkowski}

\def\Minks {Minkowski spacetime}

\def\frl{Friedmann-Lema\^ \i tre}

\def\FrL{Friedmann-Lema\^ \i tre}

\def\Godel {G\"odel}
\def\mor { $\rightarrow$  }

\def\eg {{e.g.}}
\def\ie {{i.e.}}

\def\bydef{by definition}

 \def\gr {general relativity}

\def\sr {special relativity}
\def\fb {fiber bundle}
\def\R{{\rm I\!R}}

 \newcommand{\blit}[1] {{  \emph{  #1}}}

\title{In search of relativistic time}
\author{\mlr\\
APC - Astroparticule et Cosmologie (UMR 7164)\\ ÊUniversite Paris 7 Denis Diderot
}
\begin{document}
\maketitle

\begin{center} to appear in \\ {\sl Studies in History and Philosophy of Modern Physics}\\ 
\end{center}

\abstract{This paper explores the status of   some notions which are usually associated to time, like datations, chronology, durations, causality, cosmic time  and time functions in the Einsteinian relativistic theories. It shows how,  even if some of these notions do  exist in the theory or for some   particular solution of it, they appear   usually in mutual conflict: they cannot  be  synthesized    coherently, and this is  interpreted as  the impossibility to    construct  a common entity which could be called \emph{time}. This contrasts with  the case in Newtonian physics where such a  synthesis precisely constitutes   Newtonian time.  \\
 After an    illustration by  comparing   the status of  time in Einsteinian physics with that of the vertical direction in Newtonian physics, I will conclude  that there is no pertinent notion of time in Einsteinian theories.     }

\section*{Keywords} time, causal structure, proper duration, cosmic time,  spacetime

\section{The vanishing of time}
The goal of this paper  is  to  develop and to justify  the affirmation that Einsteinian relativity is the theory of the vanishing  of time;  more exactly, of the replacement of space and time by  \spt. 
Quoting  Hermann Minkowski (1864 - 1909),  \guill The views of space and time which I wish to lay down for you have sprung from the soil of experimental physics, and therein lies their strength.
They are radical. Henceforth space by itself, and time by itself, are doomed to fade away into mere shadows, and only a kind of union of the two will preserve an independent reality.~"
(80th Assembly of the German Natural Scientists and Physicians, K\"oln, september   21, 1908).
This is true already for the \emph{special} theory of relativity, and this is its central property. This is still more crucial for the \emph{general} theory of relativity.

The analysis presented here offers  no original idea  (excepted the analogy in section \ref{analogy}). Everything  can  be found in manuals or in some
epistemological analyses concerning  the Einsteinian relativistic  theories.\footnote{ As it is well known, the Galilean or Newtonian physics is also \emph{relativistic}, in the sense that it obeys the Relativity Principle. I will however conform to the usage of calling \guill special relativity " and \guill  general relativity " the two Einsteinian theories. Hereafter, \guill Einsteinian" will refer to both theories where the status of the notions analyzed here are analogous. For instance, Einsteinian \spt ~refers both to the \Minks ~of \sr, and to the \spt ~of \gr.} 
However,    this  literature  is in general not concerned about the status of the  notions analyzed here, and about the status of time in Einsteinian physics (see however the papers of Rovelli and Anderson in the references, whose conceptions  agree with those presented here although with a different focus, more concerned with the attempts to quantize gravity; and the work of Barbour \cite{Barbour}, who defends an original position which goes however beyond the frame of Einsteinian physics). The goal here is to offer a synthesis   and a  panorama of the notions which could be claimed to be linked to time, and to give a clear explanation of what is meant by the disappearance   of time in Einsteinian theories. 

This clarification appears desirable since one  finds  often in the literature  references to \emph{time} in the analysis of a relativistic situation. This     short way to refer to a time function  or a  proper duration  may give the false impression that  such  notions have  the   properties that we usually attribute  to time. Apart from  being   a possible source of false statements, this may orient an ontological analysis into erroneous directions. 
Thus I estimate that this review may be  especially useful for philosophers. 

The paper analyzes the  notions 
of  datation, chronology, durations,  causality and cosmic time, both in  the Newtonian and  Einsteinian  theories. In Newtonian physics, these notions can be synthesized to form  \guill Newtonian time ". Here, I   defend the conclusion  that, even if some of these notions do  exist in Einsteinian  theories, or   in some    particular solution of them,   it is impossible to synthesize them   coherently to construct  a common entity which could be called \emph{time}, as    is the case in Newtonian physics.

In section \ref{Newtonian time},  I analyze  the collection of the  main characteristics of       Newtonian   time,
 in relation with the notions mentioned above, whose    synthesis   precisely leads to the construction of time (through the property of datation). 
Then, section \ref{Special relativity}    introduces \sr, and illustrates the disappearance of time there; first by showing    that some   of the above notions with  temporal flavor    have no counterpart; secondly  by  showing  that   the  different remaining  notions   --- causal structure, proper durations and  time-functions ---  cannot be    synthesized coherently to construct    an entity with the properties of time.
In section \ref{analogy} I  suggest 
an original    historical analogy  to illustrate 
 this absorption of time into \spt.  Section \ref{Relativistic theories}  introduces  the general theory of relativity and,  concentrating  on the differences with the \emph{special} theory, I show how the disappearance  of  time is still more crucial. I present (section \ref{Time functions}) 
   a specific analysis  of time-functions in  the theory  that I illustrate (section \ref{Cosmictime}) with the case of cosmology and cosmic time.

\section{Newtonian time}\label{Newtonian time}

 \guill~What, then, is time? If no one asks me, I know what it is. If I wish to explain it to him who asks me, I do not know. Yet I say with confidence that I know that if nothing passed away, there would be no past time; and if nothing were still coming, there would be no future time; and if there were nothing at all, there would be no present time. "\\ {Augustine,  (Chapter XV:17)}\\

The success of   Newtonian physics was largely due to the introduction of a frame for Physics: the Universe, unique and unified \bydef.  This    is  a prerequisite for  the universality of the physical laws,  which is itself a requirement  for the scientific methodology; in  other words, a necessary  condition for physics: no  physics without   universal law (a prerequisite for experimental methodology) ; no  universal law without an universal framework.

Newton defined this universal framework    to be composed  of   space and time, whose properties are given in his  \emph{Principia}.  In many aspects, the Newtonian conceptions about space and time  agree with those of  our intuition. This helps a lot for understanding them. On the other hand, this may be an obstacle for realizing  that they do not represent the reality of the world; in particular that there is no analog to the   Newtonian time in nature,   as well  expressed    in the Einsteinian  relativistic theories.  

Since Newton, many observational and experimental results have shown to  physicists that the real world is incompatible with the existence of Newtonian time. 
Although this is  well known, I claim that it    is incompatible with the notion of \emph{time} itself,    defined as  an entity admitting  a minimum collection of    properties that we usually attribute to time.
 Although this is well taken into account by the Einsteinian theories of relativity, this is a property of nature which does not depend on   any physical theory or model, and which is more and more confirmed by  our   progressing     experimental and  observational exploration of    the world.
How is it  possible to reconcile this fact with our      use of  the   notion  of time   in  ordinary life ? Which time-related  notions can be defined in relativistic theory?   I will try to answer these questions.
    
Although nobody, in physics or in philosophy, has claimed to give a definitive and complete  definition of time, we know quite well     the properties of  \emph{Newtonian}  time.   
Whatever   we would like to  call \guill time " should share  at least a subsample  of  these  properties. Thus, it is a prerequisite for the study of temporal notions in any theory to examine the properties of Newtonian time.

Many  particular   circumstances offer the possibility  to define some entities with temporal flavor, \ie,    which share  such or such    property  of the  Newtonian time (\eg, \guill proper times", or \guill cosmic time"). 
I  claim that we must resist  the temptation  of  referring  to them as time, since it is impossible to  reconcile them with other properties that we also desire    associate  to time,  without generating   contradictory statements. This is what I interpret as   a  manifestation of the impossibility of  existence of time.

\subsubsection{ The Newtonian  \spt}

A large part of this paper will compare some properties of Newtonian Physics with those of Einsteinian   Physics. 
But it may be  not  easy to really appreciate the deep differences between the two theories, because  they are  masked by their different formulations. For that reason, it     may appear convenient to express them in a common language, and to   reformulate part of the Newtonian approach  in order to make  easier the  comparison with Einsteinian physics. This does  not change the physics, but only the way to express it, which may appear   quite unfamiliar. Concretely, I will introduce a notion of a \emph{Newtonian \spt}, to compare it with the Einsteinian \spt. 

Both theories consider    \spt ~as a geometric entity (thus a set with geometrical properties); the occurrence of a physical event  is identified with an element (or point) of \spt.~\footnote{ It is tempting to define \spt ~as the set of all events and some modern  approaches 
adopt indeed this position. In our present vision of the world (Newtonian or Einsteinian), we  admit   the possibility   of  points of \spt ~(in Newtonian physics, a date and a location)  where   no physical event occurs; but since   an event could occur (or have occurred) there,    \spt ~may be seen as the set of \emph{possible occurrences} of events.}
This  definition    makes sense both   in the  Newtonian and in the Einsteinian  framework.  Their respective \spt s admit different       well defined geometrical  structures,  that I  will not discuss here in details.
The    main property of Newtonian \spt ~(not shared by the  Einsteinian one)    is   that  it can be \emph{uniquely}   decomposed  into a product of Space and a time-line. \footnote{ More correctly, it is defined  as  a \emph{\fb} manifold, with the time line as basis manifold.}  This splits  the Newtonian chrono-geometry into  a spatial geometry plus a chronometry. The existence of Newtonian time is equivalent to a metric structure (chronometry) of the time line, entirely independent from the  spatial geometry.
The introduction of the Einsteinian  theories results from the 
realization (first by Einstein) that    the real world phenomena are  incompatible with this conception. 

\subsection{A datation is a time-function}

The  first, and probably the more fundamental property of Newtonian time,  is \emph{datation}.  As I will show, 
the whole collection of its   other   properties    result from datation, and  I will  adopt the position   that a minimal requirement for a notion  of  \guill time" is  to be -- at least -- a   datation.
 
A datation  is a process which assigns to any  event  a real number called  its    date.
This is  a  function from \spt ~to  the set $\R$ of real numbers,  here called  the   time-line, what we    equally     call  a \blit{time-function}. 
The Newtonian datation is build-in the geometrical (\fb) structure of  Newtonian theory,  where it appears precisely as a projection of 
\spt~(with \fb ~structure) onto its basis, the time-line (the possibility of such a projection disappears in Einsteinian theories).

The datation   specifies immediately   a  unique  and well defined  total ordering of the events:  the \emph{chronological ordering} (more simply, the chronology) of \spt.

We will examine below  the possible existence of time-functions in   Einsteinian  theories. 
A first important result is that some   \spt s  wich are  solutions of   \gr ~do not admit any    time-function   (some solutions,  like   the \Godel ~\spt \cite{Godel}, admit   \emph{local} but not \emph{global} time-functions).
As a second result, a \spt ~solution which admits a time-function does admit an   infinite collection of them  (even in the simplest case   of   \Minks). 
One may wish to adopt a  particular one (like for example the  \emph{cosmic time} in cosmology, see below),  and to  promote it to the status of \guill time ". We will see that this leads  to three kinds of problems:\\
- this may be simply not be possible in  some solutions of \gr; \\
- when possible, such a choice  is  arbitrary.  Different  choices  lead   to different estimations   of  the date of an event, of  the \guill  duration " separating  two events, and even (in some cases), of  their  chronological ordering; this means that,  for two given  events A and B (not  causally connected, see below), 
one choice implies that      " A is chronologically  posterior to B " ; and the other that   "  B is posterior to A ".  \\
- the chronological notions derived from a time-function  usually  contradict physical measurements: 
physical events with different dates may appear simultaneous (and vice versa);   the   \guill  chronological duration  " of a process, as  derived from a time-function     differs in  general from its \emph{proper duration},    obtained as the result of a physical measurement.  

   We will see in which particular circumstances it is possible, in the Einsteinian context,  to chose  a time function  with a  special status,   and to  which extent it  may share  some properties of the (Newtonian) time.
  
 \subsection{The properties of   Newtonian time}

The datation defines a  \blit{chronology} :    a total ordering of all points  of   \spt, thus of all events,  thanks to their dates.  The chronological past (present / future)  of an given event is  well defined as   the subset of events with a smaller (equal / greater)  date. \\
The \blit{causal structure}
is defined as the relation which answers the question  \guill which  events are a possible    cause of   (can have an influence on) a given  event~?~ ".  The causal structure of Newtonian physics is  strictly equivalent to its chronology, thus a total ordering. \\
\blit{Simultaneity} : Events are (chronologically) simultaneous when they have the same date. \\ \blit{Space} (at time $t$)  is the subset of all  events (or of points of \spt) which share the same date $ t$.\\
\blit{Duration} is a number which can be assigned to any  physical process.  It is  a physical quantity, that we can  measure with a clock.
 
\subsubsection{Durations}

I insist on that property since its status  will be completely modified   in Einsteinian  theories. 
A duration  is an  assignment of  a real number -- its  duration --   to any process (that we also call a history).

In Newtonian physics,   duration is defined  from the    datation : the duration of a  process is the difference between the  dates of its final and initial   events. 
Hence, the  important property:  \\
(D) = the duration of a process only depends on its initial and final events. 

The Einsteinian   theories admit  the    notion  of  \emph{proper duration} of a process: a  unique  real number, assigned to it by the metric of \spt. But  it is not linked to any time function (or time) and  does not obey the   property (D)  above: it 
 depends not only on the terminal events of the process,  but on the details of its  whole history as illustrated by the   famous  twin \guill paradox ".~\footnote{ originally proposed by Paul Langevin. It   appears, however, as a paradox only when one tries to reconcile it with a notion of time.}  
  In particular,    when a \emph{time function}  is defined, the proper duration of a physical process  is \emph{not}  the differences between its values (the \guill dates ") at the terminal events.  
  
 \section{Special relativity} \label{Special relativity} 

The best characterization of the relativistic revolution 
is  the replacement of  space +   time (which together form   what we have called   Newtonian \spt)  by a 4-dimensional \spt ~as the frame for physics. 
Although Newtonian space and time each admit a separate metric,  Einsteinian  space-time admits an unique  spatio-temporal metric: this  is a 4-dimensional Lorentzian manifold, \ie, a differentiable manifold with a Lorentzian metric (see below). Any notion with temporal flavor is defined from the \spt ~metric, not from a temporal metric.

The \spt ~of special relativity  is    \Minks, the only  4-dimensional flat  Lorentzian manifold.~\footnote{ up to topological variants  \cite{LALU}.}
It is  homogeneous and isotropic, which means  that it admits   a maximal group of isometries. \footnote{In a [pseudo-]Riemanian manifold, an  isometry is a transformation (diffeomorphism) which preserves the metric. Any [pseudo-]Riemanian manifold admits a group of isometries. When it is maximal, the manifold is said to be  homogeneous and  isotropic}. Its  isotropy  implies the complete equivalence of    all    time-like  directions. {In a  Riemanian manifold (like Newtonian space), isotropy means the equivalence of all possible directions, in the sense that they can be exchanged by a rotation without modifying any property. In a Lorentzian manifold, this is slightly different since the metric divides the possible directions into three families : space-like, time-like and light-like, see below.  Isotropy implies the equivalence of all time-like directions. Newtonian \spt ~is \emph{not} isotropic since it admits a  unique time-line which has a special status compared to all other directions}

The choice of a time function (\ie,  of a datation)  selects    an arbitrary   direction in  \spt ~and breaks   this isotropy (conversely, the selection of a temporal direction  defines an infinite family of datations).   The  infinite number of different possibilities  to accomplish   such a breaking leads to an infinite number of possible time functions,     leading to mutual  contradictory statements about  temporal notions (see below).

\subsection {A Newtonian analogy}\label{analogy}

The isotropy of \Mink ~\spt ~is a fundamental fact of relativistic physics, and the deep expression of the non existence of time. 
For this reason, I will present  an  analogy\footnote{~that I have also developed   in \cite{disparition}.} between 
\emph{the  status of time in \Minks} 
and  \emph{the status of the  vertical in Newtonian space}: an analogy   between  
 \emph{the transition  from Aristotelician Physics to  Newtonian Physics}Ê 
 and \ \emph{the transition  from   Newtonian Physics to  Einsteinian Physics}.
 
\subsubsection{From Aristotle  to Newton}
\begin{itemize} 
    \item    Pre-Newtonian  physics distinguishes         horizontal and  vertical directions.         Horizontal directions form a 2-dimensional isotropic plane; 
        the-1 dimensional  vertical direction is exterior to it.\\
This is for instance manifest in the  Aristotelian claim that  the natural motion (in the sublunar world) is  vertical. 
This may be called a fundamental \blit{anisotropy} of  pre-Newtonian space.
\footnote{ although, strictly speaking,  the notion of \emph{space} is an anachronism in this context.}
      \item
The    Newtonian transition  is the replacement \\ horizontal (plane) + vertical \mor   \blit{3-dimensional isotropic space}. \\
This isotropy means that all  dimensions have the same status.
This goes against the empirical evidence since our main experience is terrestrial: we have no chance to  experience the isotropy  of Newtonian  space since it is masked by  the gravitational field of the Earth.  One aspect of the Newtonian revolution was precisely to disentangle  gravitational effects from spatial  geometry. 
   \item
All rotations are  permitted in Newtonian physics,   including those mixing vertical and horizontal dimensions.
   \item  There is no way (except in  the local terrestrial environment) to select  a  vertical among all the spatial directions. 
The   vertical appears as the signature  of  something new, and  exterior to geometry  of space:   the  local gravitational field.
   \item
There is no meaning  of the vertical in empty space: it can be any direction and there is no canonical way to select one.     
   \item In a group theoretical formulation, the   rotation group   SO(2) of 2-dimensional plane  is improved to the  rotation group SO(3) of 3-dimensional space.
    The choice of a particular direction (\eg, to be called the vertical) is a breaking of the corresponding  SO(3) symmetry; or, equivalently,  a group reduction  SO(3) \mor     SO(2).  
\end{itemize}

\subsubsection{From Newton  to Einstein}

Replacing \guill  horizontal / vertical "  by \guill  space / time "; 
and \guill  space " by \guill \spt ~"; everything can be imported to depict the Einsteinian revolution. 

\begin{itemize}
  \item  Pre-Einsteinian  Physics distinguishes  space and time. Space  forms a 3-dimensional isotropic manifold;         the one-dimensional time-line is exterior to it. This may be called the  fundamental \blit{anisotropy} of  pre-Einsteinian \spt.
      \item
The     Einsteinian  transition (\sr) :\\ space + time \mor   \blit{4-dimensional isotropic \spt}. \\
This isotropy means that all time-like dimensions have the same status.

Again,   this goes against the empirical evidence since our main experience distinguishes space and time. We have no chance to experience the isotropy  of \spt, since it is masked by our impossibility to reach high velocities (compared to $c$). Einsteinian theories  account  of this   as a consequence of our particular status, being unable to suffer high acceleration and to reach high velocities.  
    \item
All rotations are  permitted in Einsteinian   physics,   including those mixing spatial  and temporal  dimensions: they are called \emph{boosts} (or inertial  transformations).
   \item There is no way   (except in special conditions to be examined)
 to select  a \guill  time " among all the (time-like) directions. 
  \item 
In empty  \spt, any time-like direction can be chosen as a substitute of time; there is no way to select one.  One may wish to   interpret  this as \guill Time is nowhere ", or as \guill  Time is any direction in \spt ~", see below.
 \item
In a group theory formulation, the 
  rotation group   SO(3) of 3-dimensional space   is improved to the  rotation group SO(1,3) of 4-dimensional \Mink ~\spt, the Lorentz group.   
  
     The choice of a particular direction (implied by that of   a time function) is a  breaking of the corresponding  Lorentz  symmetry; or, equivalently,  a group reduction  SO(3,1) \mor   SO(3), from  the  Lorentz group to the spatial rotation group.
\end{itemize}

\subsubsection{Selection: vertical and time functions}

In empty space,     all  directions are equivalent. How  to select one? Where is the vertical ? Maybe I would like to refer to  the direction to my  (distant) planet  Earth ? Or  to  the nearest planet, or star, of which I am feeling the gravitational attraction  ? Or, more simply,  the direction to my   space ship ? Or that  of my feet  ? ...

In any case, a  choice involves   something exterior to the geometry;  something \blit{in addition} to space, usually linked to   a gravitational force.
  
   The situation is exactly similar for the direction of   time in \spt:   there are infinitely many  equivalent possible   directions which may be chosen as  time-functions. They lead to contradictory notions of, \eg, chronology. In any case,  something exterior to the chrono-geometry is needed to select  a direction: like the presence of some material system. The problem is that   different matter components define different and incompatible \guill time-functions ".
 The choice,  in Einsteinian \spt,  of a time direction which coincides with my proper time  along my own world-line (see below) has the same status than the choice, in Newtonian space,  of the vertical as the direction to my feet.
 
\section{Einsteinian Relativistic theories}\label{Relativistic theories}

An Einsteinian  theory, and more generally a metric theory, considers that the frame of physics is a \spt : : 
a   4-dimensional Lorentzian manifold, \ie, a differentiable  manifold 
with a  Lorentzian metric (and  some  additional properties). To this  metric -- which is a tensor --  is associated a 
curvature (also a   tensor)  which  expresses the\guill  shape " of \spt. 
  \begin{itemize}
  \item  In  \emph{\sr}, the metric  is entirely specified and has no curvature (it is flat). It is called the \Mink ~Êmetric, and the \spt ~is called the \Mink ~\spt. Gravitation is not taken into account.   \\ 

  \item 
 In  \emph{\gr},  the Lorentzian metric   has a  curvature  identified  with the gravitational field.  It is not fixed a priori  but is obtained \guill  dynamically " as a result of the field equations: the Einstein equations.
A  basic problem  is to find the metric of \spt, for a given configuration of matter-energy,  by solving the \Eequs ~(or another equation in some   other metric theory). 
\end{itemize}

\subsubsection{The metric}

By definition, a metric  is a tool which assigns a quadratic-interval (QI) to any   curve segment  (rigorously, to any vector).
We are used to Euclidean (or Riemanian) metrics, where the  always positive  QI  leads to the usual definition of length (as its square root). 
But the relativistic theories use a \blit{Lorentzian} metric,  which assigns a QI of  any sign. 
 
This leads first,  in both Einsteinian theories,  to a classification of the  curves\footnote{~A curve is of a given type when its tangent vector always remains of that  type.}:
they  are called time-like, light-like or space-like when their  QI is positive, zero, or negative (the opposite convention also exists). 
This is  at the origin of the causal structure, see below.
 
 This leads also  to the  definition of   the \emph{proper duration}   of a  segment of a time-like curve: this  is the square root of its  QI. \footnote{~Similarly,   the \emph{proper length} of  a space-like curve segment  is the square root of minus its QI.}
Proper duration is an attribute of a curve segment provided  by  the \spt ~metric. It  is not linked to any time,  datation or time-function.

\subsection{Time related notions in GR }

Time and space are   absent in the vocabulary of  relativistic theories; as well as  velocities. \footnote{ \emph{Four-velocities} are  however well defined in \spt, and  usually simply called \guill  velocities ".} The metric of \spt ~leads however  to the definition of various quantities and structures   with a temporal connotation. 
\begin{itemize}
\item  A  causal structure (see next section)  is   a well defined   causal ordering in \spt, imprinted by the metric.  
  \item  
A \emph{proper duration} is  assigned   to each segment of a time-like curve; in other words, to any history.
  \item  The time-related notion of \emph{redshift} is well defined and very useful in relativistic theories.  \item  
A \spt ~may or may not admit time functions,  defined in section~\ref{Time functions}. When it does, it admits an    infinite   number of them. Each may share  \emph{some} properties of Newtonian time: it  defines a datation, an associated chronology,   an associated  simultaneity...  But these notions can be defined in an infinite number of different manners, which    contradict each other.     
\end{itemize}
Contrary to the case of Newtonian physics, these notions are not mutually compatible. For instance,  \begin{itemize}
  \item    the chronology  defined from one   datation differs from the chronology   defined by another; and it    does not coincide with the causal ordering.
    \item In  two \spt s admitting  \emph{the same} causal structure, the proper durations assigned to the same history  are (in general) different. 
    \item  Given a time function, the proper duration of a history is not the difference between the terminal and initial dates of the history.      
\end{itemize}
\subsection{The causal structure }

 Any (relativistic)  \spt ~admits a   well defined     \blit{causal structure}. It is defined  from    its Lorentzian metric  
 \footnote{~in fact from its \blit{conformal part} only. Two different metrics are said to be \emph{conformally related} where one is equal to the other multiplied by a scalar function;  a conformal structure is an equivalence class under this relation.}  through the following steps:\begin{itemize}
  \item 
The Lorentzian metric assigns (\bydef) a quadratic interval (QI) to any   curve segment (rigorously, to any vector). 
   \item  This  classifies the segments as time-like, light-like or space-like according to the sign of their QI (positive, zero or negative respectively). 
 \footnote{~A conformal structure does not associate a QI to a segment, but only a sign. This is sufficient to establish the present classification.}  
  \item  By extension, this also classifies the curves; only the curves all of whose  segments (all tangent vectors) keep the same sign are retained as physically significant. 
 \item Non space-like curves are called \emph{causal}.\\
\end{itemize}

This establishes the possible causal relations between   two  events A and B  (points in \spt): \begin{itemize}
  \item 
   A   causally precedes B ($A\le B$) if  there is a  future-directed \footnote{  \spt ~is assumed to be oriented, and time-oriented.} causal curve from   A to B  (and reciprocally); 
  \item       A and B are causally disconnected, if there is no  causal curve joining them. 
\end{itemize}

This   order relation  is   the  causal structure of \spt.   A main difference with  the  Newtonian case --- where the third possibility is absent --- is that the 
ordering  is  \emph{partial}   rather than total.
An immediate  consequence is the impossibility  to  associate to an event  something which can be called  its    \emph{present}.

The causal structure is   well defined in any \spt. It is  completely  independent   from  any  notion of temporal character, like  time-function, datation,  chronology or proper durations...

It may be important to remark  that the presence of a well defined causal structure -- which is always guaranteed in any \spt ~-- does not forbid what are  called \guill   chronology    violations " (also   sometimes  \guill causality  violations ");  like, \eg,  time travel.  In fact,  \emph{time travel} is  possible in a given \spt, only if no (global)  \tf ~exists.  As a consequence, time travel is incompatible with the existence of  time. The purpose of  \Godel, when he exhibited his  solutions of \gr ~with the possibility of time travel, was precisely to illustrate  the impossibility of the existence of time
\cite{Cassou-Nogues,Godel,WangGodel,Earman,Arntzenius}.   

 
In \gr, the causal structure is defined from the metric. Some present speculative approaches try to define \spt ~without a metric, but retain the   causal structure  only. Then, it is     intrinsically defined  as an order relation in the set of events, without any reference to a metric. \footnote{ thus, without the possibility to consider proper durations, proper times, or time functions.}  This is for instance  at the basis of the \emph{causet}  (for \guill causal set")  approaches (see, \eg, \cite{Bombelli}, \cite{Dribus}, \cite{Sorkin}, \cite{Sorkin2}). 

\subsection{ World lines and Proper time}
\subsubsection{ The \wline }

The fundamental rule of relativistic theories is that \\
{\bf a material object (in particular an observer) follows a time-like line in \spt, called  its  \wline.} \\This is 
the line of its  successive positions in \spt.
Reciprocally,  each {time-like line} in \spt ~ may be seen as   the \wline ~of a potential observer.
More condensed, in a chrono-geometrical language, a  physical system (a particle, an observer ...) \blit{is} its own    \wline.

A physical process is a continuous succession of events experienced by a physical system like an observer \footnote{ I call \emph{observer} any physical system which is able to record something, in particular its proper time; an observer is typically a human being; but this can be any device with recording ability. Since an atom itself may have transitions resulting from an interaction with,\eg, radiation, it is sometimes   considered as an observer.}:  a    part of     his  life,    represented by a portion of his   \wline,  represents a  history. The metric  of \spt ~assigns   to any history --- a portion of a  time-like curve --- its   \blit{proper duration}.

The proper duration is the real physical quantity that an  observer (or   physical system) may experience and call a duration in the usual sense. A  physical measurement,  using any kind of clock, always gives the proper duration of the clock history (which is also mine if I am close to the clock). My  physiological time (the beat of  my  heart); my   intellectual time (\eg, the one necessary for a mental calculation) etc. are linked to my proper time.  This is a   fundamental assumption  of the  relativistic theories. Note again that this has no relation with    any kind of time-function.

A important point   is that   an    observer may experience --- or  measure ---      proper durations  along   his own  \wline ~only  (where he  stands);   the notion has absolutely no meaning    elsewhere in \spt. 
There is no way, and this has no meaning, to try to compare the 
  proper times of different observers, except  in the trivial case when they share (approximately)  the same \wline. 
 
 A fundamental  difference with \nwph ~comes from the fact that the proper duration of a  history \emph{is not fixed by its initial and final events}: it  depends on the whole corresponding trajectory (\wline) in \spt; and does not obey the property (D) mentioned above.
  Between two given events, there is however one unique \footnote{~Except in specific situations like gravitational lensing or a multi-connected universe \cite{LALU}.}  curve segment (a history) which has the \emph{longest} proper  duration : the geodesic. \footnote{~This is in exact  analogy with the case of Riemanian geometry in space: between two points of \emph{space}, there is a line with \emph{shortest} length: the geodesic; a straight line in flat Euclidean  space.}  This is a straight line   when   \spt ~is flat (\eg, \Minks).
In the  case where a datation has been defined, the proper duration is \emph{not} the difference of dates between its terminal events. 
 The 
\blit{Langevin twin \guill paradox "} is an illustration
of this situation: between the same initial and final events, the two twins have experienced different (proper) durations: their ages are different when they meet again.
More generally, between two specified events, two    observers  $\calA$ and $\calB$    experience different   proper durations associated to their different histories.

In the general case, the \emph{redshift} is  a tool allowing to    compare  the different \emph{perceptions} of a given process by two different observers. But   it would have no meaning to say that one proper time is flowing faster or slower than an other;  or that one duration is contracted or dilated \wrt the other.

\subsubsection{Proper time}
 
A given observer $\calO$ may  define a proper time  flowing along his world line only (that I also write $\calO$). After the choice of an arbitrary origin $O$ (a point on the \wline),  its value for   an event $A$ of $\calO$  is defined as the proper duration separating $A$  from $O$, counted with a sign which depends on the time-orientation.
This  proper time
  is  defined   on $\calO$ only, not  in the universe outside.  Thus it 
 does not define a datation, and provides      no chronology, for events out of $\calO$.

A  selected  observer (like \guill me ")  may    
wish  to  extend the validity  of his proper time  outside  the whole universe. He will
search for   a time-function  which, when restricted to his \wline ~$\calO$,  coincides with his proper time.  There is  an infinite number of different manners to perform such an extension: this  requirement  selects  an  infinite  sub-family among the  infinite family of possible  time functions. Two  observers  select  different sub-families (in some particular cases --- like  inertial observer in \Minks ~---  there is a   canonical way   to privilegiate one, see below).

\section{Time functions in relativistic theories}\label{Time functions}

The Newtonian time is a datation, \ie, a  time-function. Proper times or durations  do not  define   datations. 
Do the  relativistic theories admit time-functions   ?

The definition remains the same :     a function which assigns  to each event   a  number, and which increases along each future directed time-like line. \footnote{~That the \tf ~increases along the \wline ~of an observer does not imply that it is identical to  his proper time: it can be any   increasing  function of the proper time.}    This means that such a time-function  \guill flows " for each objet along its \wline. 

A time function generates a foliation of \spt: a decomposition of the type    of a product Space
$\times$ Time, the latter being identified  to  the time function. \footnote{
A time function generates an unique  foliation;   a foliation generates a family of time functions related by reparametrizations.}
The \guill  spatial sections " are  the   level hypersurfaces of the \tfunction. Thus, a first condition for the existence of a \tf ~is that the \spt ~admits such a splitting. This is not always the case, and  an   important result is that some \spt s -- which are  solutions of \gr  ~-- do not admit any    time function.\\
When    a time function does exist, there is    always an infinity of them. \footnote{~Analogy:   in the Newtonian   isotropic space, there exits an infinity of  equivalent  directions which may  be called the   vertical;   here, the infinity of arbitrary possibilities is still  much \guill  larger ".}
We   examine  below when, and how, is it possible    to make  a pertinent selection  among  this diversity.
This  is analog to   the search for a   vertical direction in the  isotropic  Newtonian space and, similarly, this requires (in general) a reference  external to the chrono-geometry of \spt. 

We insist on the fact that a time-function does not coincide, in general,    with the   proper time of an observer along his  \wline; and that    the proper duration of a process (what is measured by its clock)    differs from  the corresponding   lapse of a  time-function.

\subsection{Time functions in \sr}

The  \Minks ~admits  an infinity of possible  decompositions  under the form  of a product time $\times$ space. Each generates  an infinity of   time-functions, related by reparametrizations.

Since \Minks ~is flat (without curvature), it  seems reasonable to require that the spatial sections are flat also  (\ie, each identical (isometric)  to the Euclidean space    $\R^3$).
This is equivalent  to require that the time-lines are straight lines. 
This constraint reduces the plurality of time-functions, but  still leaves       an infinity of them, which lead to different  and contradictory  datations,  chronology, simultaneity. 

A further step may be to require that the  time function  coincides   with the proper time of an  inertial observer, along his \wline. This  reduces again   the choice; but    still   leaves   an infinity of possibilities,  corresponding  to all     possible spatial orientations  of an    observer. 
The isotropy  of \Minks ~ guaranties --  \bydef ~-- that  all these directions are equivalent: all these time-functions share exactly the same status;    none of them plays a  privileged role. 

A final  selection requires  the choice of  a particular inertial  observer. 
A  natural choice is \guill  me ", assuming that I am inertial, which is a reasonable approximation when high precision is not required. Thus  
I  may chose \emph{the} time function coinciding with \emph{my} proper time along my \wline ~(and defining flat spatial sections). This corresponds, in the context of \sr, to the choice made by    Newton, that  he called \emph{universal time}. (a more correct  appellation would be  \guill \emph{my} universal time ".) But if I am observing a galaxy, for instance, it could be  more convenient  to chose       the \guill universal time " of that galaxy. This  would define a completely different chronology, not compatible with the first one.

In any case, \\
-   a different observer, having    naturally chosen the  time function coinciding with \emph{his} proper time along his \wline, would adopt a chronology entirely different from mine: he  would     assign  different dates to the events, and even different chronological ordering. His notion of simultaneity  would contradict  mine.\\
-    the proper duration of a  process   (like the life time of a star, or of a particle...) never coincides     with the lapse of (my)  universal  time between birth and death.   Except   for the history of my own life, since  I defined the (my) universal time    for that purpose.
   
\subsection{Time functions in \gr}

The \Minks ~is only a very crude approximation to  the real  world, since it does not take gravitation into account.  
 In  \gr,  presently the best  theory to describe the world,  the situation is less simple.

First, the theory admits  solutions  where  {no  time function}   may exist    (globally \footnote{~For most of these solutions, it is however  possible  to  define   \emph{local} time functions; \ie,  valid in some limited region of \spt.  When restricted  to such a   limited region of the universe, the discussion  is similar to  that  below}).
A case is given for instance by the \Godel 's solution, that he   constructed with  the specific    purpose of  showing explicitly the impossibility to define time in \gr ~\cite{Godel,WangGodel}.
There are many other solutions of the theory in which no time function  may exist. Most of them allow the possibility   of time travel (whose   physical   pertinence is  still under discussion today) 
(see, \eg, \cite{Earman,Arntzenius,mlrbook}).

The majority of the solutions of \gr ~which have been proposed  
to describe realistic situations,  admit however  \emph{an infinity} of global time-functions. Since they lead to contradictory  datations, chronologies, and notion of simultaneity, this raises again the question of 
selecting one. The  situation is  similar to that in \sr ~presented above, although with some differences.

\subsubsection{Spatial sections with constant curvature}

In \sr, we applied a first prescription by selecting those  time functions    leading to spatial sections with zero curvature.
This is  not possible (in general)  in the curved \spt ~of \gr. A  less severe prescription  would consist  in   requiring spatial sections (level hypersurfaces)  to   verify some  symmetry  property, namely being of  constant curvature (homogeneous and  isotropic). Only a very limited    class of  solutions of the theory allows this possibility: the corresponding \spt s are said to obey the \blit{cosmological principle} (CP); this is the definition. 
The family of such solutions constitute 
 the \blit{\FrL ~models}.
 
 For a  \spt ~obeying the CP,   we may require in addition  that the \tfunction ~coincides with the proper time of an inertial observer. This still reduces the variety of \tfunction s. But I will   not go deeper   into the details of the procedure  because of the following remarks.

First, this   applies   only to the very  limited class  of  \spt s    obeying  the CP. This  is not the case of our  \emph{real} universe, where    galaxies and galaxy clusters imprint local curvature  at various places of \spt, so that   no  spatial sections with  constant curvature can be found: the    cosmological   description of  our Universe with the popular  \FrL ~models (including the big bang models)   is  a  zero order approximation only, which neglects all   the   curvature fluctuations  imprinted by the  cosmic structures.
The validity of  a  time-function selected  through the constant curvature requirement  cannot be extended beyond this zero order approximation. For realistic  cosmology,  its  limited   pertinence is   very difficult to evaluate.

Quite fortunately, another type of selection  procedure may be applied to (most of) these   models: that of  the  \blit{cosmic time} (defined below). At the zero order (\ie, for  the unperturbed  cosmological models), it  coincides with  the constant curvature  procedure above;  but   it   extends nicely   to the (perturbed)  real situation. Thus, the constant curvature procedure appears  useless
\cite{Beguin}.

\subsection{Cosmic time  }\label{Cosmictime}

Given a \spt, the  \blit{cosmic time} is a particular \tfunction ~defined in the following way: its value at an event  $X$ is the supremum of all the  proper durations  of all future directed  time-like curves ending in $X$. 

Not every   \spt ~admits a cosmic time. For instance,    \Minks ~does not, since such a supremum has an   infinite value for any event. When it is defined (taking finite values), it is by construction  strictly growing along each future-directed time-like  curve. 

Cosmic time is well defined for the expanding  \FrL ~models at the basis of  our cosmology, at least for those with a big bang (see, \eg, \cite{Beguin}),  through their metric.    For a specific class of (imaginary)  inertial observers, called \emph{comoving}, it  coincides for each  with his    proper-time   along his \wline. 
The terrestrial observer is, in first approximation, one of these   comoving inertial observers.
 The cosmic time, like any time function, defines a congruence of time lines   everywhere tangent to the gradient of the time function:  the world \emph{congruence}. One may imagine  that \spt ~is filled of (imaginary)  comoving observers following  these lines.

\subsubsection{Material cosmic time}

This suggests   an alternative option  \cite{RughZinkernagel}  to define a \guill material cosmic time" as the function which identifies with the proper time of  each (imaginary)     observer of  this congruence, along his worldline.
This is  a particular case of the general  possibility  to define  a time function from  a  material content. Strictly speaking,   this  requires however  that the corresponding matter (conveniently called a \emph{ cosmic clock})    occupies  \emph{the totallity} of the points  of \spt,  in order  that the  \wline s form a \emph{congruence} and that the  time-function is defined everywhere. 
We may wish, for instance, to consider the collection of  galaxies present in  our Universe as such a cosmic clock. 
 But the \tfunction ~would remain  undefined   at the  points where no galaxy is present and it is very difficult to define an averaging process in curved \spt ~(see more in \cite{RughZinkernagel}) to restore the lacking information.  
In addition,  galaxies are subject to proper motions (which superpose to the cosmic expansion),  so that 
their proper  times, flowing along their \wline s,  would not coincide with the cosmic time according to the first definition: for each galaxy, there would be a local redshift factor between the two \guill cosmic times " corresponding to the two definitions, to which no observation could give access.

But the main objection  comes from the impossibility to have any  observational access to material  cosmic time according to its  definition. Even if we assume that  galaxies (or other objects) fill the totality of \spt, and if we do not worry about their proper motions, there is no possibility to measure the proper time along a given \wline ~for an observer outside that \wline.  This makes this definition useless in practice. However, the formal \emph{existence} of such \guill  material  time functions ", even if they remain out of measurement possibilities, is an important fact which plays a role   for handling the \guill  problem of time " in \qgr ~(see below).

\subsubsection{Validity of cosmic time}

Coming back to the original definition, cosmic time, like any \tfunction,   defines spatial  sections (its level hypersurfaces which  foliate \spt). \footnote{~The \wline s of comoving matter are orthogonal to these sections.} For  the \FrL ~models -- which obey the CP -- these sections  have constant curvature,   so that  the cosmic time precisely  obeys the constant curvature criterion  mentioned above. Its advantage lies in the fact that its definition still holds in a \emph{perturbed} cosmological model, which describe our real universe. 

The  cosmic time applies only to a restricted set of solutions of \gr, which seem  however well  adapted to describe our whole universe.
Cosmology mostly uses  in fact an \guill averaged cosmic time "  :  defined  from a strict  \frl ~model  which is only    a  zero order approximation,  an  averaged version,  of  our real universe. An exact -- but unknown --  version of cosmic time does exist, which    would provide a convenient chronology for our real, perturbed,  Universe.

This assumes however that our Universe is well described by a so called  big bang model with an initial singularity. Most cosmologists however estimate today that this is not  the case, and that the present phase of cosmic expansion did not begin in such a singularity but may result, for instance, from a \emph{cosmic bounce} following a phase of cosmic  contraction. 
The corresponding \spt ~would admit no cosmic time, although some other cosmic time  functions (see below) may remain defined, at least in some part of \spt.

\subsubsection{Cosmic time is not a time}

Assuming that cosmic time is well defined, it  provides a useful  cosmic chronology, which  is its main advantage. For instance, the \guill age of the universe " $t_U$ is its  value  here and now (thus, the extremum of all the proper durations of all future-directed time-like curve ending here and now). This is perfectly well defined, and  implies  that no object can have an age greater than $t_U$. 
It should be realized however that \\
- the proper  duration of a cosmic process (say, the life of a star)  as  physically measured by a clock \blit{is  not}  (in general)  the difference  between the cosmic time  values of end and beginning. It may differ by a very important factor (which  tends to infinity for a very quickly moving object) .  \\
- For a non comoving observer, the cosmic time differs with  his    (physical)  proper   time --- the only that he has the capacity to measure --- along his \wline ~(also by an important factor). \\
- Different events  sharing the  same value of the  cosmic time   \blit{are not} simultaneous  according to the Einstein's synchronization procedure,   {even performed by   comoving observers} \cite{simult} (this is still worse for non comoving observers).
   

The cosmic time    has a local   physical    pertinence  for the terrestrial observer
since it coincides  with his \guill  universal time "  along his  \wline.
But this universal time is well defined from proper durations and the reference to  cosmic time is unnecessary in that purpose.
Moreover, cosmic time coincides with universal time only in the approximation that the terrestrial observer is   comoving.

The main defect of cosmic time, however, is the impossibility to have access to it directly: if we consider a cosmic event, like the explosion of a supernova in a remote galaxy,  no physical  measurement  provides   its  value. The latter may only be obtained through an indirect reconstitution from the redshift measurement. The latter suffers however two important drawbacks. First, un unknown component of the redshift is not cosmological and caused by the proper motion of the observed source. This introduces an error and the best we can do (and that is done in fact) is to assume that such errors average to zero for a statistical population of galaxies. This requires however to consider averaging processes which are known to suffer from biases. At  best, cosmic time estimations could be considered to be \guill true on average ". 
 The second drawback results from the fact that the conversion from a redshift (even assumed perfectly \guill cosmological ") to a value of cosmic time     requires  a perfect    knowledge of the cosmological model, \ie, of the shape of \spt. Given our uncertain    knowledge of the Hubble constant, of the deceleration parameter and of the  infinite list of similar cosmological  parameters, we must realize   that an assigned  value of cosmic time to an observed event is model dependent. 
 
 \subsubsection{Other cosmic time functions}
 
 The  cosmic time being not measurable, only (at best) statistically defined and model dependent, it would be wise to consider it as     an useful convention to describe the chronology of the  cosmic events rather than as physically relevant quantity, in any case certainly not a \emph{time}. For this reason,  cosmological calculations most often use other time-functions. \\
-   The \blit{conformal time} has    properties comparable to those of the cosmic time.
It does not identify with the proper times of inertial observers but it has more direct links with the causal structure of \spt ~than the cosmic time and this is the main reason of its preferred use in cosmological calculations. 
It also has the nice property (not shared by cosmic time)  that
different events  sharing the  same value of the  conformal time  are seen as simultaneous by  a   comoving  observer applying the Einstein's prescription  \cite{simult}.
It is however -- like the cosmic time --   not directly  measurable. \\
-  The (properly normalized )    \blit{scale factor} of the universe may be seen  as a convenient time function. It offers the same advantages as the cosmic time, except  that it does not coincides with the proper times of  inertial observers.  Its value $\tau$ is however directly   accessible from observations. Namely, for a source (galaxy)    observed with redshift $z$,    $\tau= \frac{k }{1+z}$, where $k$ is a normalization constant    which may be  is conveniently chosen as the present \guill age of the Universe "  $t_U$. One can also conveniently use   the logarithm of that  quantity.
    It is also linked to  the \cmb ~temperature $T$ through the relation $\tau= t_U~\frac{T_0}{T}$, where $T_0\approx 2.7 K$ is the present temperature of the cosmic radiation. 
    It also suffers from the fact that the measured redshift is not   cosmological  but has an unknown proper component.

 Thus, even in  the  simplest situation of a \spt ~obeying the CP,     different time functions may be chosen.  Each  choice  assigns \guill  dates "   to  the cosmic events, but they   contradict  each  other.
     \footnote{ For instance, in any  big bang cosmological model, the recombination takes place at a fixed value of redhift-time, namely $t_U/z_{rec}$, where    $z_{rec}\approx 1~100$. In a model without inflation, the value of the conformal time at recombination is of the same order of magnitude; in a model with inflation, it is almost infinite.  This discrepancy is a  direct   expression of the effect of   inflation (see, \eg, \cite{Baumann}).} 
What is the preferred  choice is a pure question of taste (\emph{cosmic time} seems good for popularization) but the best attitude is probably no choice at all since it  is perfectly  possible to perform any  cosmological calculation or reasoning without reference to any time function.  
Any choice (cosmic time or  other) would   remain  a convention, more or less adapted to such or such study,  and cannot pretend 
 to have the status of a physical quantity and to  offer a notion of time. 
  A  confusion of such notion with time 
  would be  a possible   cause  of mistake: a  neutron, whose
  (average) life time    is of the order of 14~minutes,  may  perfectly experience  a lapse of several years  of cosmic time;   a star living one million year may perfectly subsist during a lapse of several millions years    of cosmic time  \emph{etc  }.
 
   \section{Time and \gr}

To summarize, general relativity admits  two well defined  notions which are usually associated to time:   
the causal structure and the proper durations. This is not sufficient to define a notion similar to time, 
 which  requires at least
the existence of  a   time-function.  
 
First, the existence of  solutions of  \gr ~admitting no time-functions implies    that   the notion  of time is not generically  defined  in  the theory  and does not belong to its ontology. This has been  emphasized by many authors, in particular  by \Godel ~\cite{Cassou-Nogues}.   

On the other hand the theory  admits many solutions  sufficiently       regular to allow     time-functions. Selecting one  requires  however
the   adoption of      a particular and subjective   point of view, linked in general to a specific   observer,  with additional  assumptions like for instance that of a     specific   matter component filling the universe. 
Even at this price it remains that, even in the most favorable situations, chronological  propositions deduced  from a time function cannot be reconciled  with   physical measurements of a temporal nature,  in particular    proper durations. 

This  impossibility is   deeply rooted  in the Einsteinian theory and is  one of  its most  fundamental properties:  the clearest manifestation  of  the  impossibility  to include the notion   of time in its ontology: any philosophical     attempt to interpret     the   reality  of the world    in conformity with our present physics  must renounce  the notion of time. 

In a specific     situation, it is always possible to chose   a time function or another; for instance compatible with the proper time of an  observer, or of an observed system, or obeying  an alternative  convenient  prescription.   Some  authors like this  old-fashioned way but it is   dangerous  since 
a confusion with     \emph{time} would be a    source of errors.  
  
 For  precise  astronomical positioning,   spatial navigation or communication   (like  for   the decoding of the
      \emph{Global Positioning System}   (GPS)  signals),    there is no adapted choice of a   time function. In such situations, one may use convenient \spt ~\coord s without  any spatial or temporal character  like for instance    \emph{radar} or \emph{GPS}   \coord s      \cite{GPSColl,GPSmlr,GPSRovelli}.
 The study of the possibilities of  time travel (involving explicit solutions like  the \Godel ~Universe \cite{Godel}, with strongly  counter-intuitive implications) is also an active field in \gr. It is completely incompatible with any time function (since the possibility of time travel excludes the possibility of their existence \cite{mlrbook}) and the relevant tool is  the causal structure of \spt.
 
 Let me finish by shortly  mentioning  some aspects of    present  research in fundamental    physics with the quest    for a new theory with more unifying power;  which could, \eg, reconcile quantum and (general) relativistic physics. Such an enquiry has to face  very seriously the non existence of time. This is the case  for the search of a theory of  \emph{\qgr} ~\cite{qgrStanford}, where the so called Ê\guill problem of time "  \cite{Kiefer,Anderson,tproblem2} has (at least) two facets. First, our usual view of quantization requires a notion of time, which is absent in the relativistic context. \footnote{~An aspect of this non existence is technically expressed by the time-reparametrization invariance, a facet of the covariance of the \gr ~theory.}  This is one of the most serious difficulties for the tentative quantization of gravitation.~\cite{forget time}\footnote{~For instance, in \emph{Loop Quantum gravity} \cite{Rovellibook}, it appears as an \emph{Hamiltonian constraint}, which has  remained unsolved up to now.}     Secondly, a quantum \spt ~(if such a thing does exist;  it is precisely  the goal of \qgr ~to define it) would admit no defined metric (but only a \guill fluctuating one " in some popularized   language). Thus, even the  notions of causal structure  and proper durations (which, in \gr, are imprinted by the metric) disappear: not only   time is absent  in quantum gravity (if such a theory does  exists), but also the notions with temporal flavor that we have encountered in    \gr. 

This motivates a collection of speculative answers for filling this gap. Let us mention the \emph{relational time} \cite{relational,Barbour},  or clock time (or semi-classical time) \footnote{~Any   \guill measurement  of time "  is a correlation between some event and the indication of a clock.}, which may to  some extent play the role of a time function.  
The \emph{thermal   time} (a case of \emph{emergent time}) \cite{thtime1,thtime2,thtime3}  may provide a way to construct some analogous to proper time in the absence of a well defined classical (non quantum)  \spt ... These are   attempts  to define a physical notion with some   validity  in the context of a  future theory, from which time-related notions could emerge.

\section*{Acknowledgements} 
I thank Henrik Zinkernagel, and an anonymous referee,  for useful discussions about the manuscript.

  \end{document}